\documentclass[doublecol]{epl2} 
\usepackage{epstopdf}
\usepackage{subfigure}
\usepackage{graphicx}
\usepackage{amsmath}
\usepackage[cp1251]{inputenc}

\title{Stellar age dependence of the nonextensive magnetic braking index: a test for the open cluster $\alpha$Per}
\shorttitle{Stellar age dependence of the nonextensive magnetic braking index} %Insert here a short version of the title if it exceeds 70 characters

\author{Daniel B. de Freitas}
\shortauthor{D. B. de Freitas}

\institute{Departamento de F\'{\i}sica, Universidade Federal do Cear\'a, Caixa Postal 6030, Campus do Pici, 60455-900 Fortaleza, Cear\'a, Brazil}

\abstract{Using a generalized function of the stellar spin-down law, we investigate the age dependence of the magnetic braking index ($q$). Our survey includes 9 open clusters aged lower than 1 Gyr and ranged in mass from 0.7 to 1.1$M_{\odot}$. Our aim is to verify the time behavior of the nonextensive braking index $q$ which brings the cumulative distribution of the rotational velocities of the stars of the youngest cluster ($\alpha{\rm Per}$) taken at the future age of an older cluster. As a result, the $q$-index is calculated over time $t-t_{\alpha{\rm Per}}$, where $t$ is the age of older open cluster used to estimate the future cumulative distribution of the rotational velocity of the $\alpha{\rm Per}$ cluster with present-day age $t_{\alpha{\rm Per}}$. We find that the values of $q$ are slightly constant around 1.36 and 1.38 according to the mass bin. In conclusion, the results seem to indicate that the mechanism that controls the rotational decay of stars in open clusters does not depend on the increment of time.}

\begin{document}

\maketitle

\section{Introduction}

Star clusters are traditionally classified as galactic open clusters and globular clusters. Open clusters are generally rather young ($\leq$1 Gyr), whereas globular clusters are old ($\geq$10 Gyr), more massive, and dense \cite{fujii,pacepas}. In the present work, we focus our attention on the first group. 

Open clusters offer a unique material when compared to field stars because they share the same age, initial metallicity, and primordial gas \cite{mm}. In particular, the age of an individual star cannot be measured. We are restricted to the information that there are certain stars that are very young and others that are very old. Because all of the stars in a cluster are presumed to have begun their life at approximately the same time, we can estimate the age of a large group of stars more precisely.  Owing to this feature, Barnes \cite{barnes2003} showed that the presence of slow and fast rotator populations in open clusters could indicate distinct ``rotational sequences''. In addition, these populations are associated with the different dependencies that link angular momentum loss rates to angular velocity driven by stellar magnetic wind \cite{defreitas2014}.

It is widely accepted that magnetic braking is a fundamental concept for understanding angular momentum losses due to magnetic stellar winds for several classes of stars, such as main-sequence field and cluster stars. This mechanism was initially suggested by Schatzman \cite{Schatzman}, who pointed out that slow rotators have convective envelopes. As mentioned by Kraft \cite{Kraft}, the behaviour of the mean rotational velocity of low-mass-main-sequence stars below 1.5$M_{\odot}$ (spectral type F0) is preferentially due to magnetic wind. A few years later, Skumanich\cite{sku}'s pioneering work argued that stellar rotation, activity, and lithium abundances for solar-like stars obeys a simple relationship given by $\mathrm dJ/\mathrm dt \propto \Omega^{3}$, where $t$ is time, $J$ is the angular momentum and $\Omega$ denotes the angular velocity. The above-mentioned authors established early on that stellar rotation and age should be related to cool main-sequence stars. The next step was to develop a more complete theory capable of explaining how losses of angular momentum occur.

Inspired by \cite{mestel1968,mestel1984,mestel1987}'works, Kawaler \cite{kawaler1988} elaborated on a theoretical model for describing the behaviour of the loss of angular momentum for main-sequence stars with masses less than 1.5$M_{\odot}$ due to wind ejected by stars. This wind gets caught by the magnetic field that spins outward until it is ejected, affecting the angular momentum and causing slowdown. In this way, the magnetic field acts like a brake. As the magnetic field strength depends on stellar mass, Chaboyer \textit{et al.}\cite{chaboyer1995} modified the Kawaler's parametrization and introduced a saturation level into the angular momentum loss law. On the other hand, Krishnamurthi et al. \cite{kris1997} proposed the inclusion of a Rossby scaling at the saturation velocity for stars more massive than 0.5$M_{\odot}$. More recently, the Roosby number\footnote{This number is defined as the ratio between the rotational period and the convective overturn timescale.} is used to characterize the deviation from the Skumanich law.

Barnes \cite{barnes2003} proposed a simple formulation that uses the colour and period values to derive the stellar ages for solar- and late-type stars. According to authors, the age and colour dependence of sequences of stars allow us to identify their underlying mechanism, which appears to be primarily magnetic. This determination of stellar ages from their rotational periods and colours, he named ``Stellar Gyrochronology''. Gyrochronology is a age-dating technique that provides precise and accurate ages for low mass (G, K and M) field stars on the main-sequence \cite{gallet} and open clusters \cite{barnes2020}. The theoretical background is centred on a functional formulae based on the Skumanich-type age dependence (square root of time) fitted by the following expression\begin{equation}
\label{int1}
P=g(t)f(B-V),
\end{equation} 
where $P$, $t$, and $B-V$ are the rotational periods (days), ages (Myr), and colours, respectively. The function $g(t)\propto t^{1/2}$ is the rotation-age relation from Skumanich \cite{sku}. However, only a sequence of stars, also called sequence $I$, respects this condition. This sequence consists of stars that form a diagonal band of increasing periods with increasing $B-V$ colours in a colour period diagram. There is also another sequence of stars denoted by the letter $C$ that represents the fast rotators. For these stars, the magnetic field is expected to be saturated. Consequently, there is no clear dependence between the period and colour. In this case, the expression for the angular momentum loss rate is given by $\mathrm dJ/\mathrm dt \propto \Omega$, and therefore, the period and age are related by a simple exponential law \cite{chaboyer1995,barnes2003,barnes2007,pacepas,defreitasetal15}.

Recently, de Freitas \& De Medeiros \cite{defreitas2013} revisited the modified Kawaler parametrization proposed by Chaboyer \textit{et al.} \cite{chaboyer1995} in the light of the nonextensive statistical mechanics \cite{tsallis1988} using a generalized exponential law. In this context, de Freitas \& De Medeiros \cite{defreitas2013} analyzed the rotational evolution of the unsaturated F- and G- field stars that are limited in age and mass within the solar neighbourhood using a catalogue of $\sim$16000 stars in the main sequence \cite{holmberg2007}. More recently, de Freitas et al. \cite{defreitasetal15} generalized the Reiners \& Mothanty \cite{reiners2012}'s torque using the nonextensive framework. In both cases, they use the $q$-index extracted from the Tsallis formalism as a parameter that describes the level of magnetic braking. They also linked this parameter with the exponent of dynamo theory ($a$) and with magnetic field topology ($N$) through the relationship $q=1+\frac{4aN}{3}$ \cite{defreitasetal15}. As a result, they showed that the saturated regime can be recovered in the nonextensive context, assuming the limit $q\rightarrow1$. This limit is particularly important because it represents the thermodynamic equilibrium valid in the Boltzmannian regime. According this work, the torque in the generalized version is given as $\mathrm dJ/\mathrm dt \propto \Omega^{q}$, indicating that the rotational velocities of F- and G-type main-sequence stars decrease with age according to $t^{1/(1-q)}$. The values of $q$ obtained by de Freitas \& De Medeiros \cite{defreitas2013} suggest that it has a strong dependence on stellar mass.

\subsection{Aims and Structure of the paper}
The main goal of the research is to verify the behavior of the magnetic braking index $q$ which will bring the cumulative distribution of the rotational velocities of the cluster $X$ stars with age $t_{X}$ taken at the future age $t_{Y}$ of the cluster $Y$. In this sense, our paper is summarized as follows: working sample is presented in Sect. 2. In Sect. 3, we describe nonextensive framework and the magnetic braking law. In next section, we present the methodology and procedures to estimate the $q$-index. In Sect. 5, we apply our method to set of open clusters and present the results. In last section, we present our conclusions and final remarks.

\section{Open cluster samples}

In this study we analyzed 946 angular velocity data\footnote{Consequently, free of ambiguities arising from $\sin~i$, contrary to what occurs with a sample of the projected rotation ($v\sin i$) which depends of angle $i$ to the line of sight.} in 9 open clusters aged between 35 and $\sim$800\,Myr and ranged in mass from 0.7 to 1.1$M_{\odot}$ as presented in Table~\ref{tab1}. With these features, it is possible to investigate the effects on rotational evolution after the period of gravitational contraction indicated by the Kelvin-Helmholtz time\footnote{The Kelvin-Helmholtz timescale is the time required to radiate the current gravitational binding energy of the sun at its current luminosity. For a solar-type star, this time is of the order of 30 Myr.}. In this respect, the present paper deals with cluster stars past their zero-age main sequence (ZAMS) \cite{Silvaetal13}.

The cluster ages were obtained from the Catalogue of Open Cluster Data (COCD) by \cite{Kharchenko05}, except Hyades, whose age was obtained from the WEBDA database \cite{Mermilliod04}. Ages from the COCD were determined using the turn-off isochrone technique. The accuracy of this method is limited by the lack of high mass stars evolving off the main-sequence, particularly for young clusters. We used COCD data because it is a uniform database on cluster ages that uses a homogeneous cluster age scale. This is an important aspect, since we are analyzing the relationship between $q$ and stellar age, where cluster age is the chronometer. In order to account for the possible range of cluster ages, we used the 0.33\,dex range adopted by Silva \textit{et al.}\cite{Silvaetal13} as the possible range of individual cluster ages instead of the estimated value reported in Kharchenko \textit{et al.}\cite{Kharchenko05} (i.e., 0.2--0.25 dex). Finally, we choose to reject spectroscopic binaries, because the rotation may be altered by tidally induced synchronism as quoted by Mayor \& Mermilliod \cite{mm}.

\begin{figure}
	\begin{center}
		\includegraphics[width=0.44\textwidth,trim={2.5cm 1.5cm 2.5cm 1.5cm}]{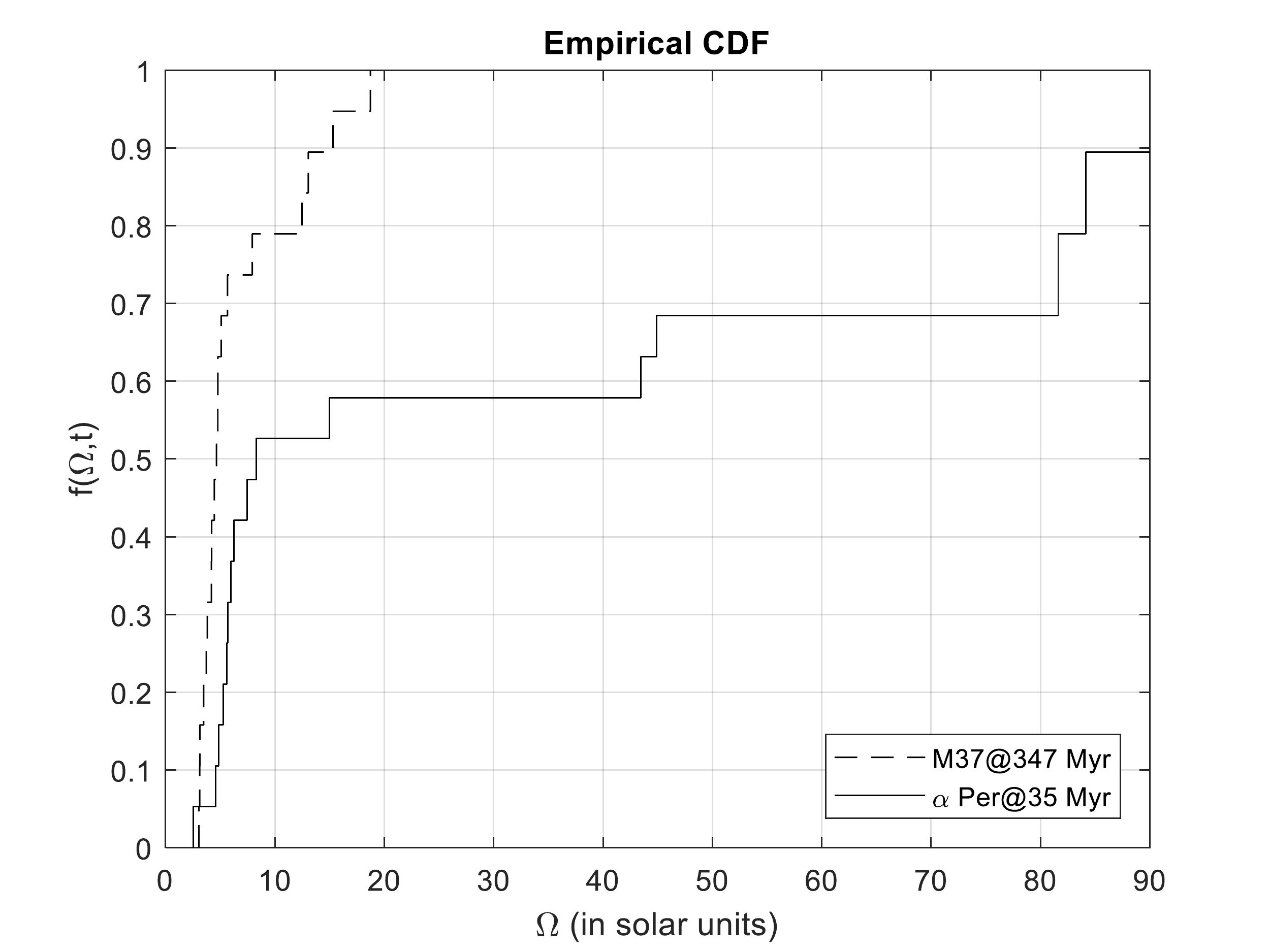}
	\end{center}
	\caption{The empirical cumulative distribution function (CDF)
of the angular velocity $\Omega$ (in solar units, $\Omega_{\odot}$) for stars
in the $\alpha$Per (solid line)
and M37 (dashed line) clusters with age of 35 and 347 Myr, respectively. The interval of mass considered is limited to $0.9<M(M_{\odot})<1.1$.}
	\label{fig1}
\end{figure}

\begin{figure}
	\begin{center}
		\includegraphics[width=0.44\textwidth,trim={2.5cm 1.5cm 2.5cm 1.5cm}]{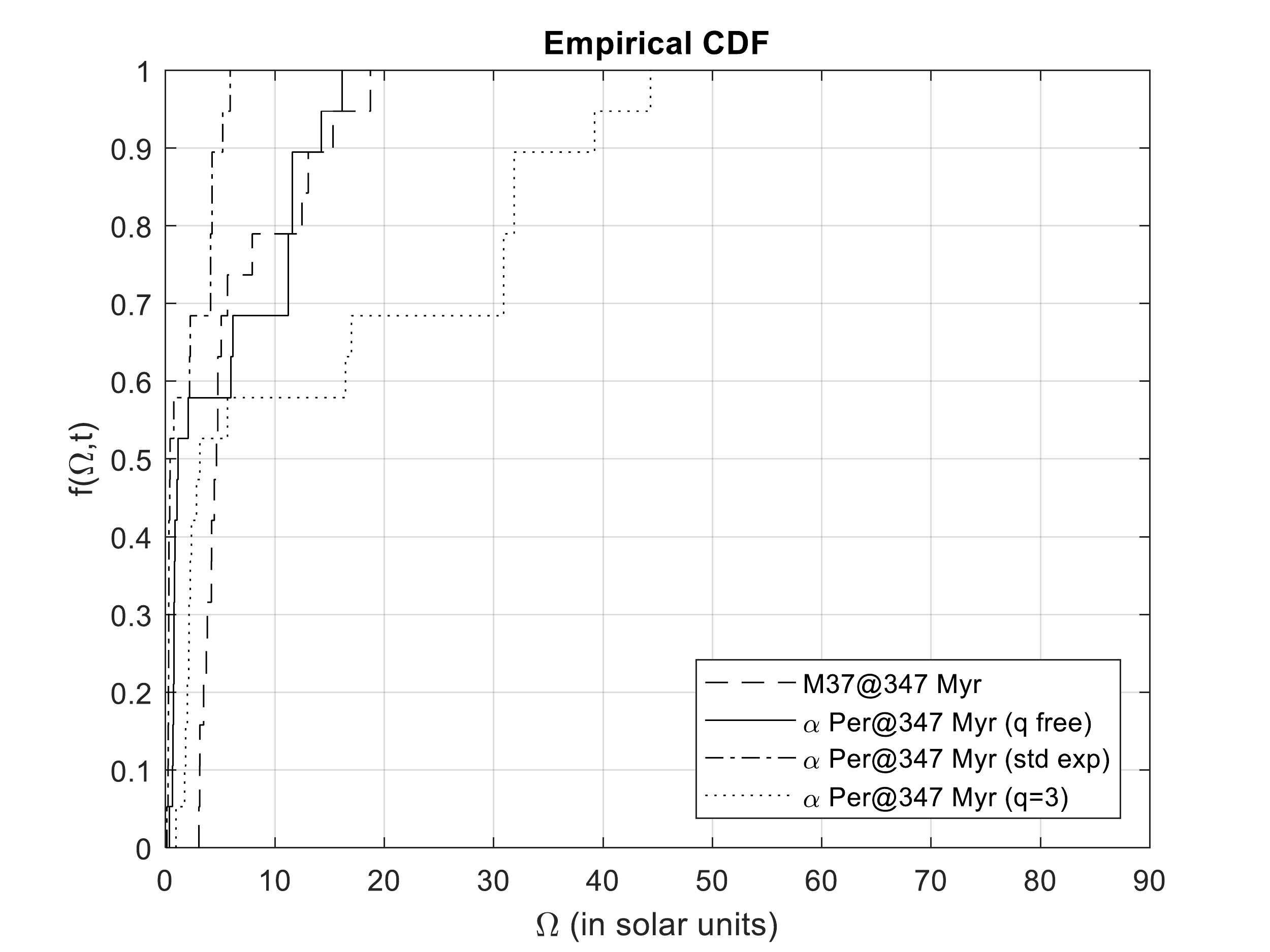}
	\end{center}
\caption{The empirical cumulative distribution functions (CDF) of the angular velocity $\Omega$ (in solar units, $\Omega_{\odot}$) for stars in the $\alpha$Per computed at age of
the M37 (dashed line), assuming the
braking index $q$ and $t_{0}$ as free parameters (solid line). The best parameters are $q=1.33$, $\tau=131$ Myr (see Table \ref{tab2}). The fits of standard exponential (dash dotted line) and Skumanich law ($q=3$, dotted line) are also shown.}
	\label{fig2}
\end{figure}

\begin{figure}
	\begin{center}
		\includegraphics[width=0.45\textwidth,trim={2.5cm 1.5cm 2.5cm 1.5cm}]{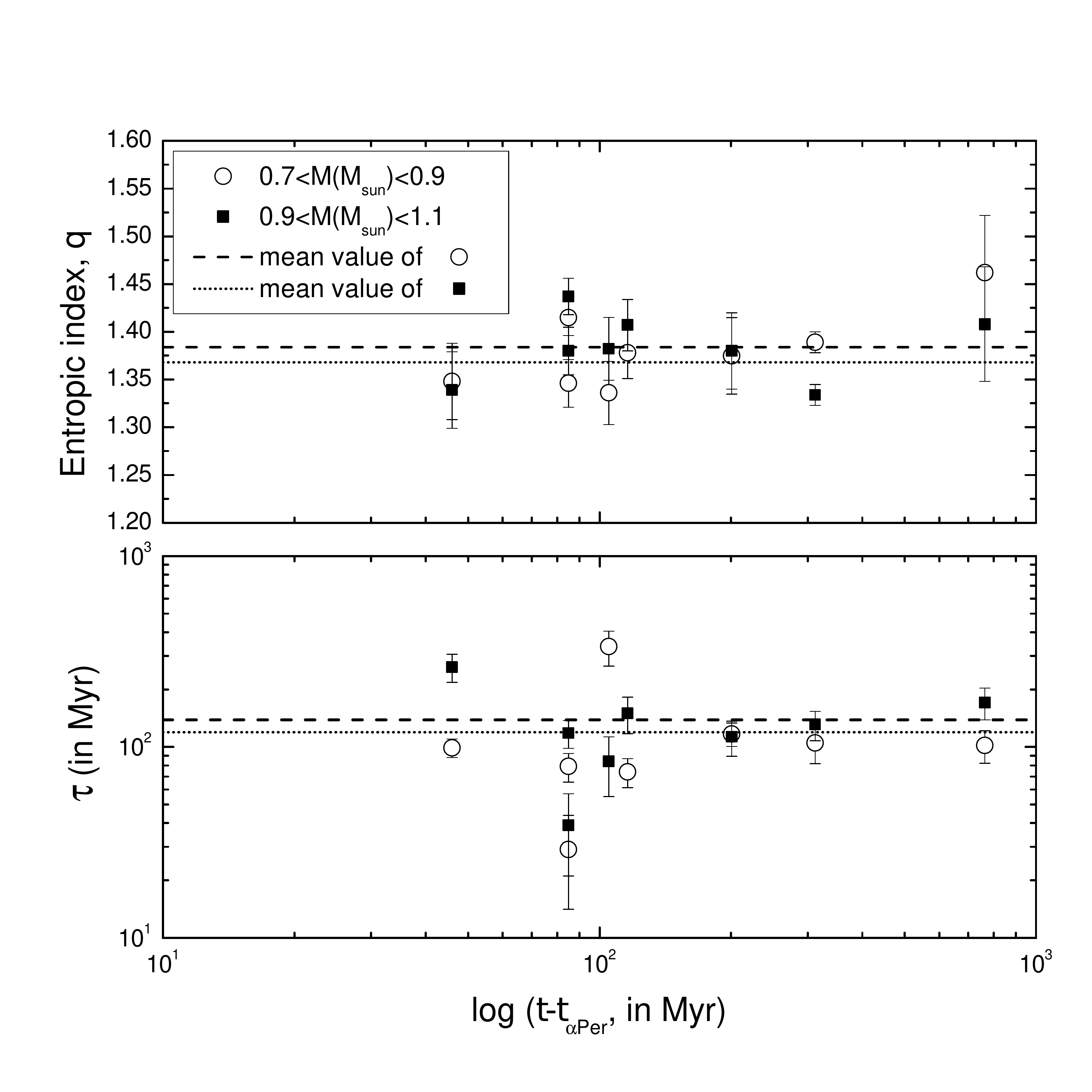}
	\end{center}
\caption{Time evolution of the $q$-index and $\tau$ computed from the best fit as shown in Fig. \ref{fig2}. The result is obtained to $\alpha{\rm Per}$ cluster in different ages based on the age of other 8 open clusters of mean age lower than 1 Gyr.}
	\label{fig3}
\end{figure}

\begin{table}
\scriptsize
\caption{Main characteristics of the data analyzed in this study limited to a range of mass between 0.7 and 1.1$M_{\odot}$. Column (1) identify the cluster. Column (2), and (3) present the references for the period, and the number of stars analyzed, respectively. Columns (4) shows the cluster ages, in Myr.}
\label{tab1}
\renewcommand{\arraystretch}{1.5}
\begin{tabular}{rc|rr}
\hline \hline
 &(2) & (3) &(4) \\
Cluster & Ref.\footnote{(1) \cite{irwin2008}; (2) \cite{Hartmanetal10}; (3) \cite{Irwinetal09}; (4) \cite{Meibometal09}.} & $N$ & Age \\
 &  &  & (Myr) \\
\hline \hline
{\bf$\alpha$Per} & 1 & 56 & 35 \\[1.6ex]
IC\,2391 & 1 & 31 & 76 \\[0.6ex]
NGC\,2516 & 1 & 41 & 120\\[0.6ex]
Pleiades & 2 & 168 & 120 \\[0.6ex]
M\,50 & 3 & 305 & 135\\[0.6ex]
M\,35 & 4 & 227 & 151\\[0.6ex]
M\,34 & 1 & 42 & 236 \\[0.6ex]
M\,37 & 1 & 60 & 347\\[0.6ex]
Hyades & 1 & 16 & 794\\[0.6ex]
\hline
\end{tabular}
\end{table}

\begin{table}
\scriptsize
\caption{Best $q$ and $\tau$–values and their errors determined using bootstrap resampling method for each pair of clusters. The reference cluster is $\alpha$Per and each value is computed by $t_{YX}=t-t_{\alpha\rm Per}$. The order of results follows the list of clusters presents in Table\ref{tab1} starting from IC2391 to Hyades.}
\label{tab2}
\renewcommand{\arraystretch}{1.5}
\begin{tabular}{rrrrr}
\hline \hline
$t-t_{\alpha\rm Per}$	&	$q$	&	$\delta q$	&	$\tau$	&	$\delta\tau$	\\
(Myr)	&		&		&	(Myr)	&	(Myr)	\\
\hline
\multicolumn{5}{c}{$0.7<M(M_{\odot})<0.9$}\\
\hline
41	&	1.35	&	0.02	&	99	&	11.0	\\
85	&	1.42	&	0.03	&	29	&	14.9	\\
85	&	1.35	&	0.03	&	79	&	13.5	\\
100	&	1.34	&	0.04	&	335	&	69.8	\\
116	&	1.38	&	0.03	&	74	&	12.7	\\
201	&	1.38	&	0.03	&	117	&	16.4	\\
312	&	1.39	&	0.01	&	105	&	23.1	\\
759	&	1.46	&	0.05	&	102	&	19.9	\\
\hline
\multicolumn{5}{c}{$0.9<M(M_{\odot})<1.1$}\\
\hline
41	&	1.34	&	0.04	&	262	&	44.0	\\
85	&	1.44	&	0.02	&	39	&	17.9	\\
85	&	1.38	&	0.03	&	118	&	19.5	\\
100	&	1.38	&	0.03	&	84	&	29.0	\\
116	&	1.41	&	0.03	&	150	&	32.7	\\
201	&	1.38	&	0.04	&	113	&	23.4	\\
312	&	1.33	&	0.01	&	131	&	23.1	\\
759	&	1.41	&	0.06	&	171	&	32.6	\\

\hline
\end{tabular}
\end{table}

\section{Statistical framework}
de Freitas \& de Medeiros \cite{defreitas2013}, using a wide sample, showed that the high values of $q$ extracted from radial velocity distributions reveals effects of long-range interactions consistent with the $q$-CLT (non-extensive central limit theorem). Our choice comes from the observational evidence that astrophysical systems are somehow related to nonextensive behavior \cite{silva,liu,viana,gell,defreitas2012}. As mentioned by de Freitas et al.\cite{defreitas2014}, it is known that gravitational systems (e.g., open clusters) with strong long-range interactions can not properly be described by usual Boltzmann-Gibbs statistical mechanics (hereafter BG).

Several authors (e.g., \cite{Tsallis1,Tsallis3,Tsallis4}) have shown that nonextensive formalism is a powerful statistical tool for studying systems
out of equilibrium with long-ranged interactions and long-ranged
memories as is observed in open clusters. This typical group of stars can be studied as a non-linear dynamic system and, therefore, the complexity of dynamics is far beyond the Boltzmannian system. Thus, it is better described by nonextensive Tsallis statistics based on the extended concept of $q$-entropy
\begin{eqnarray}
\label{eq1}
S_{q}=\frac{k}{q-1}\left(1-\sum_{i=1}^{\rm W}p_{i}^{q}\right),
\end{eqnarray}
where $k$ is a positive constant, $p_{i}$ denotes the probability for
occupation of $i$-th state, and $W$ is the total number of the
configurations of the system. In particular, when $q\rightarrow1$, the extensive BG entropy is recovered. In addition, eq.~\ref{eq1} can be maximized as defined in \cite{Tsallis4}.

In the present study, we use the parametrized approach of the magnetic braking law proposed by Mayor \& Mermilliod \cite{mm} and de Freitas \& De Medeiros \cite{defreitas2013} formally written as the non-linear
equation
\begin{eqnarray}
\label{4}
\frac{\rm d \Omega}{\rm d t}=-\frac{\Omega_{0}}{\tau}\left(\frac{\Omega}{\Omega_{0}}\right)^{q}, \quad q\geq 1
\end{eqnarray}
where $\Omega_{0}$ is the angular velocity at time $t=0$ and $\Omega$ is the velocity at time $t=t_{\rm age}$, i.e., now, and $q$ as mentioned by de Freitas \& De Medeiros \cite{defreitas2013}. Owing the negative sign presents in eq. (\ref{4}), its solution is the $q$-exponential function $\exp_{q}(-x)$ defined as
\begin{eqnarray}
\label{4sol}
\exp_{q}(-x)\equiv\left[1-(1-q)x\right]^{\frac{1}{1-q}},
\end{eqnarray}
where the $q$-extension of logarithmic function is the reverse of $\exp_{q}(x)$
\begin{eqnarray}
\label{4sollnq}
\ln_{q}(x)\equiv\frac{x^{1-q}-1}{1-q},
\end{eqnarray}
which reduces the $q$-entropy (see eq.~\ref{eq1}) to $S_{q}=k\langle\ln_{q}(1/p_{i})\rangle$ (cf. \cite{pavlos}).

In the astrophysical literature (e.g., \cite{defreitas2013}), the $q$-index has been limited to the range of 1 (stars with saturated magnetic field and, therefore, more massive ones) to 5 (corresponding to the Reiners \& Mohanty \cite{reiners2012}'s relation for the very low-mass stars).

On the other hand, the coefficient $\tau$ is a characteristic time that denotes the decay constant defined as a function of spectral type. Not only does the mean velocity depend on the magnetic braking law, but the shape of the $\Omega$ distribution is also very sensitive to index $q$ \cite{mm}. At a given stellar mass (or spectral type), $\tau$ is a constant over time and, therefore, the solution of eq. (\ref{4}) in terms of eq. (\ref{4sol}) is reduced to
\begin{eqnarray}
\label{eq4p}
\Omega(t)=\Omega_{0}\exp_{q}\left(-\frac{t}{\tau}\right),
\end{eqnarray}
for $1-(1-q)\frac{t}{\tau}\ge 0$ and $\Omega_{0}\ge \Omega$. The eq. (\ref{eq4p}) is normalized when $\Omega_{0}=(2-q)/\tau$ \cite{picoli}. As quoted by de Freitas \& De Medeiros \cite{defreitas2013}, $q>1$ denotes the unsaturated magnetic regime, whereas $q=1$ is the index associated to the saturated regime. Roughly speaking, $\Omega_{0}$ can be interpreted as a initial probability distribution in which it is possible to estimate the distribution of $\Omega$ at a given time. 

An important point is that eq. (\ref{eq4p}) describes the behaviour of the disc-less stars and, therefore, it is not suitable for very young stars, because they have a thick dust disc in the infrared \cite{costa}. On the other hand, accreting stars has the physics more complex than that of the non-accreting stars, because the star-disc interation controls the rotation \cite{vidotto}. This question is interesting because the presence of an accretion disc or even planets in the stars in our sample could put our results in doubt. In this way, a deeper analysis on the infrared excess due to cold non-condensed dust and the loss of angular momentum due to the planets would help to find the best model to understand the $q$-index.

\section{The rotational future of the $\alpha$Per cluster}
Based on the previous equations, our problem is to investigate the following question: which is the braking law which will bring the cumulative distribution function (CDF) of the rotational velocities of the cluster $X$ stars with age $t_{X}$ taken at the age $t_{Y}$ (where $t_{X}<t_{Y}$) in agreement with that of the cluster $Y$? In this sense, we are seeking for calibration of parameters $\tau$ and $q$ so that we can estimate the ``future'' cumulative distribution of $\Omega$ for a given star cluster only knowing its present-day angular velocity distribution and the age of an older cluster.

According to eq. (\ref{4}), the ``future'' angular velocity distribution of the cluster $X$ stars at a time $t_{Y}$, considering the initial distribution $\Omega_{X}$ at time $t_{X}$, is given by
\begin{eqnarray}
\label{a}
\Omega_{X}(t_{Y})= \Omega_{X}(t_{X})\exp_{q}\left(-\frac{t_{YX}}{\tau}\right),
\end{eqnarray}
where $t_{YX}=t_{Y}-t_{X}$, in Myr. Based on the Table \ref{tab1}, $t_{X}$ is given by age of $\alpha$Per cluster and $t_{Y}$ denotes the age of older open cluster than it, accordingly $t_{YX}=t-t_{\alpha\rm Per}$. We chose $\alpha$Per simply because it is the youngest cluster and, therefore, we can explore the entire time span of our sample, i.e., from 35 to $\sim$ 800 Myr.

In summary, we have adopted the following procedures: Firstly, let $f(\Omega,t)$ be the CDF of angular velocity at a given age $t$. We shall assume that this distribution is the same for any cluster at the age $t$ at a given mass bin. Two distributions $f(\Omega_{X},t_{X})$ and $f(\Omega_{Y},t_{Y})$ are known and are described in Section Open Cluster Samples. In this context, it is expected that cumulative distributions $f(\Omega_{X},t_{Y})$ and $f(\Omega_{Y},t_{Y})$ are statistically similar. Basically, that is to say, for instance, that the Hyades are the Pleiades in the future!

Secondly, it is to perform a $t$-test of two independent distributions to estimate the value of $q$ required for the $\alpha{\rm Per}$ cluster to have a cumulative distribution similar to the distribution of an older cluster. In this sense, we want to evaluate whether a correlation exists between $X$ and $Y$ cluster data. To that end, the significance of correlation coefficient can be estimate using a $t$-statistic. Thus, we specify the null and alternative hypotheses: (i) the null hypothesis $H_{0}:\rho=0$ (there is no association) and (ii) the alternative hypothesis $H_{A}:\rho\neq0$ (a nonzero correlation could exist) for the two-tailed test or $H_{A}:\rho<0$ (a negative correlation could exist) and $H_{A}:\rho>0$ (a positive correlation could exist) as the results for the left- and right-tailed tests, respectively. 

Third, we calculate the value of the $t$-statistic using the following equation:
\begin{equation}
\label{eq11a}
t_{\rm calculated}=\frac{\overline{X}-\overline{Y}}{\sqrt{\frac{\sigma^{2}_{X}}{n}+\frac{\sigma^{2}_{Y}}{m}}},
\end{equation}
where $\overline{X}$ and $\overline{Y}$ are the sample means, $n$ and $m$ are the distribution sizes, $\sigma^{2}_{X}$ and $\sigma^{2}_{Y}$  are the standard deviations of each CDF. 

Fourth, we use a $t$-table to find the critical value ($t_{critical}$), considering a 95\% confidence level and, consequently, a significance level $\alpha$ equal to 0.05 and 0.025 for each tail related to one and two sided tests, respectively \cite{trauth}. In hypothesis testing, a critical value is a point on the $t$-test distribution that is compared to the calculated $t$-statistic to determine whether the null hypothesis is rejected or not. 

Latter, we compare the calculated $t$-statistic (see eq.~\ref{eq11a}) to the critical value. In general, if the absolute value of the calculated $t$-statistic is greater than the critical value, then the null hypothesis can be rejected at the 95\% level of confidence in favor of the alternative hypothesis \cite{trauth,press}.

\section{Results and Discussions}
We have analyzed the behavior of the $q$-index, obtained from empirical CDF of the angular velocity as a function of stellar mass and age. Table \ref{tab2} summarizes the results from such an analysis, from where one can observe that the values of $q$ are systematically greater than 1, irrespective of the stellar parameter considered, in reinforcing that the Tsallis exponential function fits the observed stellar angular velocity better than the standard exponential function and the Skumanich law ($q=3$) as can be seen in Fig.~\ref{fig3}. 

All $q$-exponentials are two-parameters ($q$--$\tau$) nonlinear functions given by eq. (\ref{eq4p}). The result of our calculations for these two parameters, for well defined stellar-mass intervals, are presented in Table \ref{tab2}. Since the values of $q$ are significantly different from 1, this result strongly suggests that the distribution of angular velocity for the present sample of open cluster stars is far from being in agreement with a standard exponential, independent of the stellar mass considered. In addition, the error bars in Fig.~\ref{fig3} correspond to a 0.05 confidence limit and are shown in Table \ref{tab2}.

Our null hypothesis admits that the correlation between $X$ and $Y$ distributions is zero. For the left-tailed test, the $t$-statistic reveals that, because $\rvert t_{\rm calculated}\rvert>\rvert t_{\rm critical}\rvert$ for all the sets of samples in our study, we can reject the null hypothesis in favor of the alternative hypothesis; i.e., there is a strong correlation between each pair of distributions. Similarly, for the two-tailed test, the $t$-statistic reveals that because $t_{\rm calculated}$ is outside the range between $-t_{critical}$ and $+t_{critical}$, we also can reject the null hypothesis in favor of the alternative hypothesis. This means that the CDF of $\alpha{\rm Per}$ cluster at same age of older one (e.g., IC2391 or NGC2516) generated after optimizing the $q$-index is very similar to the CDF of older cluster. 

Carvalho et al. \cite{carvalho} showed that for stellar clusters older than 1 Gyr exists a linear trend between the parameter $q$ of radial velocities distributions and the stellar age, while for young clusters ($<$1 Gyr) this correlation is small or can be neglected. Even considering the different stellar parameters studied, this result for young open clusters is also observed in our work. Interestingly, the great majority of the young clusters analyzed by Carvalho et al. \cite{carvalho} are located in the solar neighborhood near the galactic disc, namely at the same region where are located the stars of the present sample. From Fig. \ref{fig3}, the values of the $q$ are roughly constant over the time and are consistent with the average value of 1.38 and 1.36 for two mass regimes indicated in figure, respectively. Table~\ref{tab2} and  Figure~\ref{fig3} also demonstrate that there are two regimes where the decay constant value $\tau$ is slightly constant ($>$120 Myr) and a scatter domain for values less than 120 Myr. The average values of $q$ and $\tau$ are also indicated in the referred figure for each mass bin.

\section{Concluding remarks}

We used the $q$-exponential function as a generalized one of the stellar spin-down law to investigate the behavior of angular velocity of a sample of 946 open cluster stars of mass ranging between 0.7 and 1.1M$_{\odot}$ with ages lower than 1 Gyr. In addition, we investigate the age dependence of the magnetic braking index ($q$) from the cumulative distribution of the angular velocities of the stars of the youngest cluster ($\alpha{\rm Per}$) taken at the future age of an older cluster. 

Our results showed that the $q$-index calculated over time $t-t_{\alpha{\rm Per}}$ is slightly constant around 1.37$\pm$0.01. In this context, the results seem to indicate that the mechanism that controls the rotational decay of stars in open clusters does not depend on the increment of time. On the other hand, the values of $q$ greater than 1 reveal the effects of long-range interactions and the formation of high-energy tails consistent with the $q$-CLT, where the nonextensive framework is observed.

Finally, the strongest conclusion in this study concerns the role of the $q$-index as an important parameter to explain the magnetic braking mechanism that controls the evolution of stellar rotation of open cluster stars. We conclude that $q$ is a parameter insensitive to the stellar mass and the magnetized wind that controls these stars in the main-sequence phase. In addition, we suggest that the magnetic braking index ($q$) can be an interesting and alternative way to better understand the idea of stellar rotation as an astronomical clock. In this context, future missions, such as PLATO, will be ideally suited to derive accurate stellar ages (for both cluster and field stars) and will allow us to improve magnetic braking models and age-rotation relationship.

\acknowledgments
DBdeF acknowledges financial support 
from the Brazilian agency CNPq-PQ2 (rant No. 311578/2018-7). Research activities of STELLAR TEAM of Federal University of Cear\'a are supported by continuous grants from the Brazilian agency CNPq.

\end{document}